\documentclass[conference]{IEEEtran}
\usepackage[pdftex]{graphicx}
\usepackage[usenames,dvipsnames,table,xcdraw]{xcolor}
\usepackage{array}
\usepackage{multirow}
\usepackage[normalem]{ulem}
\useunder{\uline}{\ul}{}
\usepackage{booktabs}
\usepackage{supertabular,booktabs}
\usepackage{longtable}
\usepackage{lscape}
\usepackage{cite}
\usepackage{url}
\usepackage{hyperref}
\usepackage{dirtytalk}
\usepackage{graphicx}
\usepackage{comment}
\usepackage{makecell}
\usepackage{svg}
\usepackage{amsmath}

\usepackage[caption=false]{subfig}
\graphicspath{{Images/}}
\usepackage[normalem]{ulem}
\useunder{\uline}{\ul}{}

\usepackage{tabularx}
\usepackage{balance}
\usepackage{nameref}

\begin{document}

\title{Remote Communication Trends Among Developers and Testers in Post-Pandemic Work Environments}

\author{

\IEEEauthorblockN{Felipe Jansen}
\IEEEauthorblockA{Dell Technologies\\
 Petrolina, PE, Brazil \\
 felipe.jansen@dell.com}
\and

\IEEEauthorblockN{Ronnie de Souza Santos}
\IEEEauthorblockA{University of Calgary\\
Calgary, AB, Canada \\
ronnie.desouzasantos@ucalgary.ca} 

}


\IEEEtitleabstractindextext{%
\begin{abstract}
The rapid adoption of remote and hybrid work models in response to the COVID-19 pandemic has brought significant changes to communication and coordination within software development teams, affecting how various activities are executed. Nowadays, these changes are shaping the new post-pandemic environments and continue to impact software teams. In this context, our study explores the characteristics and challenges of remote communication between software developers and software testers. We investigated how these professionals have adapted to the unique circumstances imposed by COVID-19, especially because many of them have now become permanent in the software industry. In this process, we explored their communication practices and interaction dynamics and how they potentially affect software evolution and quality. Our findings reveal that the transition to remote and hybrid work has resulted in notable changes in communication patterns and task coordination, which could potentially affect the overall quality of project deliverables. Additionally, we highlight the importance of adapting existing workflows, introducing new management practices, and investing in technology to facilitate remote interaction among developers and testers.
\end{abstract}

}

\maketitle

\IEEEdisplaynontitleabstractindextext

\IEEEpeerreviewmaketitle

\section{Introduction}
\label{sec:introduction}

The global spread of COVID-19 in early 2020 triggered a significant transformation in the working landscape, with remote work becoming essential across numerous sectors of the economy \cite{bick2023work, boland2020reimagining}. Particularly within the software industry, several software professionals found themselves forced to quickly adapt to this reality, often facing challenges resulting from the sudden transition to environments that were not designed primarily for work activities, resulting in adverse working conditions and significant levels of stress \cite{ralph2020pandemic, de2022grounded, neto2021deep}. However, these professionals adapted to the circumstances over time, leading to remote work structures (including hybrid work) becoming more prevalent in software development than ever before. Yet, challenges still persist \cite{butler2021challenges, aagren2022agile, de2023post}. 

For example, in software development, effective communication is essential for achieving desired outcomes \cite{alzoubi2014agile, treude2011effective}. The success of a software project heavily relies on the efficient dissemination of both general information and specific agreements about requirements, technical decisions, and planning, whether directed at specific team members or shared with the entire group \cite{kluender2017team, pikkarainen2008impact}. The effectiveness of communication channels is particularly evident in different development settings; in agile development, for instance, direct and simplified communication, "face-to-face" interactions, and daily meetings are essential elements that could be limited in remote environments \cite{aagren2022agile, dorairaj2011effective}.

In fact, aspects of communication and coordination have been recently discussed in many studies addressing the characteristics of the post-pandemic reality and the remote work structures in software engineering \cite{de2022grounded, bezerra2020human, miller2021your, aagren2022agile}. These findings indicate that communication disruptions present a significant challenge in contemporary working environments involved in remote dynamics. In software development, for instance, teams that lack regular interactions might encounter coordination problems and disengagement among team members, as coordination relies on the exchange of information for interdependent tasks \cite{de2022grounded, aagren2022agile, russo2021daily}. In this context, collaboration and effective communication between programmers and testers are fundamental for the software development process \cite{cruzes2016communication, zhang2014sources}. 

Hence, considering the current landscape of the software industry, marked by a significant portion of professionals working in fully remote or hybrid arrangements worldwide, and taking into account the communication challenges highlighted in the literature, our study seeks to investigate the communication processes between developers and testers within remote environments and explore their potential impacts on software projects. To guide our study, we pose the following research question:

\smallskip \smallskip
{\narrower \noindent \textit{\textbf{RQ.} How is remote communication shaping the work of developers and testers in post-pandemic software engineering environments?} \par}
\smallskip \smallskip

Efficient communication between developers and testers is essential in the software industry \cite{cruzes2016communication,grechanik2010bridging}. While developers craft the code \cite{storey2008todo}, testers ensure its quality by identifying and reporting any bugs or issues \cite{florea2023roles}. Hence, the synergy between these professionals ensures that the software adheres to desired quality standards \cite{cruzes2016communication}. Therefore, our contributions to this theme are as follows:


\begin{itemize}
\item {An investigation within the industry setting, exploring the post-pandemic communication dynamics between software developers (i.e., programmers) and software testers.}
\item {A discussion on the characteristics and challenges of remote communication in the current remote work environment in the software industry.}
\item {Initial practices to address communication challenges and enhance collaboration between developers and testers in remote work settings.}
\end{itemize}






From this introduction, our study is organized as follows. Section \ref{sec:method}, we describe our method. In Section \ref{sec:findings}, we present our results, which are discussed in Section \ref{sec:discuss}. Finally, Section \ref{sec:conclusions} summarizes the contributions of this study.

\section{Method} \label{sec:method}
The literature reveals extensive research on the intersection of software engineering and communication \cite{alzoubi2014agile, treude2011effective, cruzes2016communication,grechanik2010bridging, de2022grounded, bezerra2020human, miller2021your, aagren2022agile}. However, in the current software industry landscape, it is important to explore the impacts of the post-pandemic work environment on the communication between developers and testers, especially considering those now working in remote setups (including hybrid work). Hence, we chose a survey-based methodology as our main research approach to understand this phenomenon from a professional and practical perspective. In this sense, we conducted a cross-sectional survey \cite{easterbrook2008selecting} following well-established guidelines in software engineering~\cite{ralph2020empirical} to explore the remote communication phenomenon. Below, we describe the details of our method. 

\subsection{Questionnaire} \label{sec:question}
The survey instrument was designed to explore diverse perspectives on communication processes within remote software development teams, considering the unique dynamics shaped by the COVID-19 pandemic. However, our primary focus remained on investigating the impact of communication between developers and testers in this environment. This approach aimed to foster the comprehension of the intricate nature of communication in contemporary software development environments. Following the formulation of the questions, the questionnaire underwent a pilot validation process through three consensus meetings involving researchers and software professionals. During these sessions, suggestions for refining, adding, or removing questions were carefully considered—the finalized questionnaire comprised five open-ended inquiries, six closed-ended queries, and nine demographic questions. Access to the completed questionnaire is available here\footnote{https://figshare.com/s/14a610752b34081d2e5d}.

\subsection{Participants} \label{sec:participants}
The surveyed population comprises software professionals working in coding and quality activities (e.g., developers, software engineers, tester engineers, test analysts, and testing managers, among others) who, after the pandemic, remained working in remote and hybrid environments. These professionals originate from various software organizations across different industries, including but not limited to mobile, manufacturing, finance, healthcare, e-commerce, education, telecommunications, and software development by demand, and they engage with the development of software applications across diverse contexts and methodologies. This diversity offers a broad spectrum of experiences and viewpoints on the remote communication dynamics of software development. In our study, we consider remote environments, those where individuals work on tasks performed outside traditional offices but within the same geographic area, which differs from the traditional concept of global software development that requires collaboration across different countries \cite{de2022grounded}. Additionally, while acknowledging the fundamental differences between remote and hybrid work, our study primarily focuses on communication processes, in particular, on the one main characteristic: \textit{being remote}. We highlight that in both remote and hybrid scenarios, team members commonly work outside the office, either intermittently or permanently, and need to interact remotely at some level. This shared trait defines our concept of remote communication.

\subsection{Data Collection} \label{sec:collect}
We followed the recommendations for sampling in software engineering \cite{baltes2022sampling} and employed two data collection techniques: convenience sampling and snowball sampling, both of which are examples of non-probabilistic sampling. Initially, we sampled software professionals known to have experience in remote environments. These individuals were selected from the authors' extensive network of industry professionals, especially those who had participated in previous studies on remote and hybrid work. Subsequently, the snowballing technique enabled us to expand our participant pool beyond this initial network, as participants were requested to share the questionnaire with their peers, thus facilitating further recruitment. Data collection occurred during mid-2023. 

\subsection{Data Analysis} 
\label{sec:analysis}

Initially, descriptive statistics methods~\cite{george2018descriptive} were applied to describe the key characteristics of our sample. Through statistical functions such as means, proportions, totals, and ratios, we categorized participants' responses into sub-groups, providing valuable insights into the dataset. Subsequently, thematic analysis~\cite{cruzes2011recommended} was conducted to explore the responses from participants to open-ended questions. This method facilitated the identification of recurring themes and patterns within the qualitative data, thereby enriching our understanding of participants' experiences and perspectives. 


\subsection{Ethics} 
\label{sec:ethics}
Adhering to ethical guidelines, this study refrained from collecting any personal information about the participants, such as names, email addresses, or employers, to uphold their anonymity. 

\section{Findings} 
\label{sec:findings}
Our sample consists of 154 professionals engaged in both development and quality activities, providing us with a variety of perspectives and experiences within the field. We also obtained details on the frequency at which developers and testing professionals meet to discuss aspects related to the project and the online tools used to support this interaction. Finally, we have evidence of the characteristics of this communication, challenges, and initial practices to address the potential problems (Tables \ref{tab:charac}, \ref{tab:challenges}, \ref{tab:address}). We expect that these results can support practitioners in improving the communication among developers and testers, which can ultimately improve several quality activities, including software verification and validation, maintenance and evolution.

\subsection{Demographics} \label{sec:demo}
In summary, the sample primarily consisted of developers (82\%) and male participants (72\%). In terms of professional experience, senior respondents (with over ten years of experience) accounted for 61\% of the total participants, followed by mid-level professionals (6 to 9 years of experience) at 20\%. Moreover, our sample comprises highly qualified individuals, with 35\% holding educational qualifications beyond an undergraduate degree, including postbaccalaureate certificates and master's degrees. Geographically, 51\% of respondents were from Brazil, with Indian participants comprising the second-largest group at 26\%, followed by Americans at 7\%. Responses were also collected from professionals in China, Germany, the United Kingdom, Canada, Israel, Mexico, Ireland, Malaysia, and Portugal (one respondent from each country). In addition to different geographic backgrounds, our sample included representatives of various underrepresented groups in software engineering, with 48\% identifying as non-white, 22\% as female, and 5\% as part of the LGBTQIA+ community.

\subsection{Remote Communication Configuration} \label{sec:meetings}
Our findings reveal that programmers and testers have embraced a remote communication configuration that could support robust collaborative practices, with a majority of participants frequently meeting to make project decisions. This collaborative inclination is evident among 76 participants who reported constant interaction with peers during the day (e.g., through meetings and chats), followed by 52 participants who reported occasional interactions (e.g., predetermined meetings or for specific purposes).

The trend towards maintaining an open communication channel reflects a mutual appreciation for maintaining synchrony among development and testing activities, enhancing collaboration within the work environment. However, in contrast, we identified a total of 26 professionals (13 developers and 13 QAs) who reported a communication configuration based on limited interactions with their peers (e.g., mostly when there was a major decision-making process in progress). This scenario may suggest either ineffective communication among these professionals or a preference for greater independence and autonomy, which the new work environments (e.g., remote and hybrid) can naturally accommodate.

Regarding tools supporting remote communication, 117 participants reported using Microsoft Teams for remote communication, citing its user-friendly interface, seamless integration with other tools, and alignment with organizational policies as primary reasons for its choice. Additionally, 52 participants utilize Slack, while 41 opt for WhatsApp, both valued for their versatility in addressing various communication needs, especially synchronous interaction via chat, supporting personal and small-scale project interactions. Furthermore, 23 participants mentioned Google Chat as a secondary option for communication in specific contexts. Another secondary tool identified by 17 participants was phone calls, particularly in instances of internet connectivity issues. Notably, we observed limited adoption of enterprise tools, indicating potential gaps in their effectiveness or acceptance within the sampled professionals. While the tool list may not be groundbreaking, these findings underscore the importance of understanding professionals' preferences and the efficacy of these tools in promoting effective remote communication between development and testing.

\subsection{Remote Communication Characteristics} \label{sec:characteristics}
Developers and testers have seamlessly integrated remote communication into their workflow. However, they emphasize that for remote communication to be effective, it should support several forms of \textit{synchronous} interactions (e.g., meetings, chats, live discussions), especially during important discussions, such as those related to feature planning, bug reporting, and debugging. Conversely, asynchronous communication methods were perceived as \textit{lagging}, particularly within agile environments. Furthermore, participants highlighted that remote communication could be \textit{unreliable}, citing external factors such as intermittent internet access, power outages, or audio disruptions as occasional challenges. Concerns about communication quality were also identified among respondents, leading to the necessity for additional meetings to ensure effective mutual understanding.

Additionally, they highlighted that remote communication can be \textit{limiting}, particularly for tasks that require extensive interaction, such as assisting new team members. However, this limitation is often attributed to the effectiveness of online tools rather than the concept of remote communication itself. However, despite potential challenges associated with remote communication, professionals generally maintain a positive perspective on its benefits. They perceive remote communication as more \textit{empowering} than in-person interactions, emphasizing its flexibility, autonomy, and ability to accommodate diverse professional needs and preferences. This approach supports individuality and allows each person to tailor their communication style to suit their unique requirements.

\begin{table}
  \caption{Remote Communication Characteristics}
  \label{tab:charac}

\begin{tabularx}{\linewidth}{p{1.8cm} X}
\toprule
Characteristic & Evidence Examples \\ \hline 

Synchronous & ``I believe it has improved significantly; I feel much closer to the leadership team now.'' (P33) \newline ``This has increased the availability of team members, making it much easier to address questions or concerns about specific issues, whether through text messages or voice communication.''(P52) \newline \\

Lagging & ``Sometimes we ask something and the response takes time, and vice versa.'' (P29) \newline ``The communication flows smoothly overall, although occasional hiccups happen, such as individuals being unavailable to communicate or delays in receiving responses or feedback.''(P37) \newline \\

Unreliable & ``Remote communication does not affect my work. There are external factors that sometimes make communication difficult, such as lack of internet, power outages, or noise.'' (P11) \newline ``Lack of network and electricity.''(P88) \newline \\

Limiting & ``There are an effort of the team to response as fast as possible, but yes, some simple points and misunderstandings always happen often than personally.'' (P116) \newline ``For new developers, it's challenging to assist them because online interaction becomes somewhat limited, especially considering the tools currently available in the company..''(P25) \newline \\

Empowering & ``Discover the solution by myself most of the time before approaching someone for help.'' (P140) \newline ``It allows each person to work during their most productive hours of the day, actually having a positive impact.'' (P48) \\

\bottomrule
\end{tabularx}

\end{table}

\subsection{Remote Communication Challenges} \label{sec:challenges}
Remote communication, now widespread in software organizations, brings various challenges and complexities in coordinating development and test activities. One critical factor is \textit{impersonality in interactions}, which affects communication dynamics. The lack of face-to-face interactions creates an atmosphere some describe as monotonous, making effective \textit{collaboration} difficult. The absence of non-verbal discussions, crucial for building trust and understanding, might hinder the collaboration among developers and testers.

Another significant challenge resulting from remote communication is the possibility of \textit{reduced team engagement}, which can impact \textit{productivity}. Without the spontaneous interactions and immediate availability of colleagues that occur in a traditional office, team members often find it difficult to get quick responses and timely support. This delay can slow down decision-making processes and create bottlenecks, especially in fast-paced environments where quick problem resolution is crucial. Additionally, the absence of physical proximity limits opportunities for brainstorming and informal knowledge sharing about the code and the tests.

Furthermore, remote communication causes \textit{lack of clarity}, represented as the difficulty in fully understanding project details, which significantly affects the \textit{coordination} between developers and testers. This challenge often leads to an increase in the number of meetings as team members attempt to compensate for incomplete or unclear information. Moreover, excessive dependence on written communication can occasionally result in misunderstandings or overlooked messages, adding complexity to the coordination between coding and testing. In this scenario, while some teams may manage to maintain effective coordination despite these hurdles, others may find themselves struggling to ensure that developers and testers are on the same page. 

\begin{table}
  \caption{Challenges in Remote Communication}
  \label{tab:challenges}

\begin{tabularx}{\linewidth}{p{1.8cm} X}
\toprule
Challenges & Evidence Examples \\ \hline 

Impersonally & ``Not really affecting any deliverables in remote conversation but we are missing the personal contact with team members.'' (R143) \newline ``The impersonality (chat, meeting), I believe, is what most affects communication.'' (P13) \newline \\

Reduced Engagement & ``Although we are able to efficiently perform assigned tasks and activities, there is a diminishing effect per say.'' (R149) \newline ``Reduce communication efficiency. Someone will not work hard.'' (R153) \newline \\

Lack of Clarity & ``Every small discussion is translating into a meeting now. Knowledge sharing is becoming difficult.'' (R86) \newline ``I believe the biggest challenge is conveying accurate and complete information. Sometimes, out of fear of omitting details, many unnecessary meetings are scheduled, often at the last minute. This results in an excess of meetings, leaving insufficient time to fulfill other commitments.'' (P06) \\

\bottomrule
\end{tabularx}

\end{table}

\subsection{Practices to Address Remote Communication Challenges} \label{sec:solutions}

Professionals are addressing the challenges by implementing various solutions, with one key strategy: development and testing integration. One approach involves maintaining a direct communication channel among developers and testers, enabling early identification of issues before finalizing development tasks. This involves professionals empowered to take action to improve communication schemes to address potential issues proactively. Teams have also implemented pre- and post-completion overviews of tasks. This includes regular alignment meetings to address questions, align understanding, showcase task completion, demonstrate the implementation process, and maintain standardized information sources. Finally, mandatory communication channels involving the entire team, including developers, testers, architects, designers, and managers, ensure effective communication not only within coding and testing activities but across the entire project.

\begin{table}
  \caption{Addressing Challenges in Remote Communication}
  \label{tab:address}

\begin{tabularx}{\linewidth}{p{1.8cm} X}
\toprule
Strategies & Evidence Examples \\ \hline 

Direct Communication & ``Certain important tasks got missed while communicating through teams. So we had to take a stand to email certain important things so they can be tracked effectively.'' (R105) \newline ``Misunderstandings about technical details happened in the past. By talking to each other again, we were able to clarify the requirements''. (R14) \newline \\

Proactivity & ``With many people needing to discuss the same issue in a production problem, we created a war room to resolve the problem in parallel.'' (R13) \newline ``At times, delays in receiving responses from the client via chat impacted our delivery cadence. To address this issue, we began sending emails with managers copied in to ensure everyone was aware of the need for faster communication.''(P39) \newline \\

Regular Alignment & ``The test/QA analyst is integrated into the development team, working together. This way, we can catch many issues before finalizing the development.'' (P44) \newline ``Because everybody is remote, hence sometimes responses can be delayed. However, with multiple follow up through e-mails and discussion, that is no longer an issue.'' (P115) \\
\bottomrule
\end{tabularx}

\end{table}

\section{Discussions} \label{sec:discuss}
Considering the varying degrees of communication frequency reported in our sample, remote communication in software engineering operates along a spectrum, ranging from constant interaction to limited communication focused on essential project decisions. This spectrum seems to be responsible for the intensity to which developers and testers perceive the impact of the challenges associated with this work structure, which potentially impact collaboration, coordination, and productivity at different levels. Therefore, a key distinction between the pre-pandemic working environment and now is the increased possibility that individuals and teams have in configuring this communication spectrum, allowing for greater control over communication channels, interactions, and management of challenges and their solutions. Our analysis highlights the critical importance of clarity and focus in remote communication, with an emphasis on minimizing unnecessary meetings and addressing specific project-related issues. While delays in responses can pose risks to project quality, effective communication strategies, hybrid work policies, and adaptable practices can help mitigate these challenges.

Therefore, in answering our research question, this variation shapes how developers and testers perceive and handle challenges related to collaboration, coordination, and productivity. Post-pandemic environments allow individuals and teams to configure their communication approaches more easily, allowing for more control over interactions and problem-solving.

\subsection{Implications} \label{sec:implications}
This study provides important insights for the software industry, highlighting the need for companies to adapt workflows and invest in strategies that enhance communication between programmers and testers in remote and hybrid environments. By focusing on effective tools and improved interaction practices, companies can boost collaboration, coordination, and productivity. Moreover, while the study is industry-focused, it also makes a valuable academic contribution by filling a gap in the existing literature on remote work in software engineering. By offering a unique perspective on the dynamics between development and testing, particularly in the post-pandemic context, our research expands the understanding of software teamwork in this new era.

\subsection{Threats to Validity} \label{sec:limitations}
While this study provides valuable insights into the dynamics of remote communication among software developers and testers, several limitations and threats to validity must be acknowledged. First, the findings are derived from the experiences of a sample of 154 professionals, which may limit the generalizability of the results. However, the lessons learned from this study may still be transferable to various contexts within different software companies. Additionally, the data collection method relied on self-reported survey responses, which could introduce response bias or result in incomplete information. Examples of potentially ambiguous or incomplete information responses include statements such as "Does not negatively affect," "Affects positively," "Quite productive," "Not at all," "Do not believe it affects," "Nothing Much," and "Not so much," which may lack precision.

Moreover, the study did not examine the impact of time zone differences on team communication. While this is an important factor, it falls outside the scope of this research, as the focus was on remote and hybrid scenarios where team members typically work outside the office post-pandemic, either intermittently or permanently, but remain within the same country and time zone, choosing not to work in person. Lastly, the rapidly evolving nature of remote work environments presents challenges in accurately capturing the most current practices and challenges. Despite these limitations, this study contributes to a deeper understanding of the nuances of remote communication in software engineering teams and offers valuable insights for both practitioners and researchers.

\section{Conclusions} 
\label{sec:conclusions}

In this study, we explored the dynamics of remote communication among developers and testers within software engineering teams in the post-pandemic era. Our goal was to understand their communication methods, challenges, and strategies, providing valuable insights for practitioners and contributing to the knowledge on remote work in software engineering. Through a survey with industry professionals, we identified insights into the multifaceted nature of remote communication dynamics, highlighting both opportunities and challenges. In summary, while remote work offers flexibility, it also presents complexities such as maintaining clear communication channels, which can impact collaboration and productivity within developers and testers. Therefore, we emphasize the importance of adopting effective communication tools, interaction practices, and protocols to ensure transparent and efficient communication in remote settings.

Moving forward, we understand that our future research endeavors should explore the evolution of this trend. For example, with several software companies requiring their professionals to return to the office, we are interested in understanding what new characteristics, challenges, and benefits this communication configuration might have on developers and testers, especially considering that recent studies have highlighted the problems of returning to the office to these activities. Additionally, we want to further explore the nuances of remote communication and their impact on software evolution, particularly in terms of technical debt management. Moreover, efforts to develop and validate frameworks for assessing and enhancing remote communication effectiveness would be invaluable in guiding organizations toward more resilient and productive remote work practices.

\ifCLASSOPTIONcaptionsoff
  \newpage
\fi

\balance
\bibliographystyle{IEEEtran}
\bibliography{bib.bib}

\end{document}